\documentclass[twocolumn,aps,showpacs,amsmath,amssymb,floatfix,superscriptaddress]{revtex4}
\usepackage{graphicx}
\usepackage{dcolumn}
\usepackage{bm}
\usepackage{hyperref} 
\usepackage{latexsym}
\usepackage{float}
\begin{document}

\title{Divergent Time Scale in Axelrod Model Dynamics}
\author{F. Vazquez}\email{fvazquez@buphy.bu.edu}
\author{S.~Redner}\email{redner@bu.edu}
\affiliation{Center for BioDynamics, Center for Polymer Studies, 
and Department of Physics, Boston University, Boston, MA, 02215}

\begin{abstract}
  
  We study the evolution of the Axelrod model for cultural diversity.  We
  consider a simple version of the model in which each individual is
  characterized by two features, each of which can assume $q$ possibilities.
  Within a mean-field description, we find a transition at a critical value
  $q_c$ between an active state of diversity and a frozen state.  For $q$
  just below $q_c$, the density of active links between interaction partners
  is non-monotonic in time and the asymptotic approach to the steady state is
  controlled by a time scale that diverges as $(q-q_c)^{-1/2}$.

\end{abstract}  


\maketitle

A basic feature of many societies is the tendency to form distinct cultural
domains even though individuals may rationally try to reach agreement with
acquaintances.  The Axelrod model provides a simple yet rich description for
this dichotomy by incorporating societal diversity and the tendency toward
consensus by local interactions \cite{A}.  In this model, each individual
carries a set of $F$ characteristic features that can assume $q$ distinct
values; for example, one's preferences for sports, for music, for food, {\it
  etc}.  In an elemental update step, a pair of interacting agents $i$ and
$j$ is selected.  If the agents do not agree on any feature, then there is no
interaction.  However, if the agents agree on at least one feature, then
another random feature is selected and one of the agents changes its
preference for this feature to agree with that of its interaction partner.  A
similar philosophy of allowing interactions only between sufficiently
compatible individuals underlies related systems, such as bounded confidence
\cite{compromise} and constrained voter-like models \cite{VR}.

\begin{figure}[ht]
 \vspace*{0.cm}
 \includegraphics*[width=0.39\textwidth]{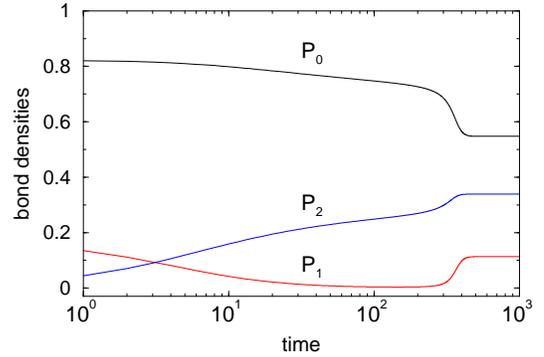}
 \caption{Master-equation time dependence of bond densities $P_0$, $P_1$, and
   $P_2$ for $q=q_c - 4^{-1}$.  Each agent has 4 neighbors.}
\label{Pm}
\end{figure}

\begin{figure}[ht]
 \vspace*{0.cm}
 \includegraphics*[width=0.39\textwidth]{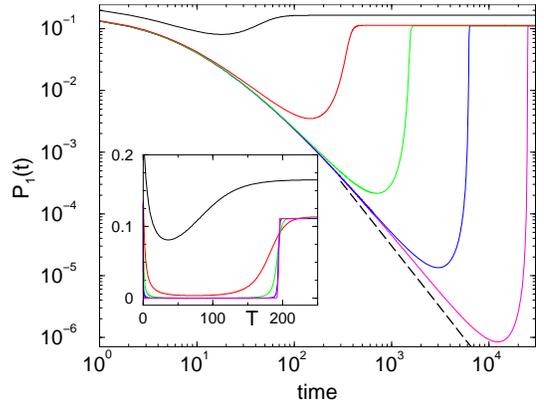}
 \caption{Master-equation result for $P_1(t)$ for $q=q_c-4^{-k}$, with $k=-1,
   1, 3, 5$, and $7$ (progressively lower minima).  Each agent has 4
   neighbors.  The dashed line has slope $-2$ (see text).  Inset: Same data
   on a linear scale with $T=t(q-q_c)^{1/2}$.}
\label{P1vst}
\end{figure}

Depending on the two parameters $F$ and $q$, a phase transition occurs
between cultural homogeneity, where all agents are in the same state, and
diversity \cite{A,CMV,VVC,KETM}.  The latter state could either be frozen,
where no pair of interacting agents shares any common feature, or it could be
continuously evolving if pairs with shared features persist.  The rich
dynamics of the model does not fall within the classical paradigms of
coarsening in an interacting spin system \cite{spins} or diffusive approach
to consensus in the voter model \cite{voter}.  In this Letter, we solve
mean-field master equations for Axelrod model dynamics and show that the
approach to the steady state is non-monotonic and extremely slow, with a
characteristic time scale that diverges as $q\to q_c$ (Figs.~\ref{Pm} \&
\ref{P1vst}).

The emergence of an anomalously long time scale is unexpected because the
underlying master equations have rates that are of the order of one.  Another
important example of wide time-scale separation occurs in HIV \cite{hiv}.
After an individual contracts the disease, there is a normal immune response
over a time scale of months, followed by a latency period that can last
beyond 10 years, during which an individual's T-cell level slowly decreases
with time.  Finally, after the T-cell level falls below a threshold value,
there is a final fatal phase that lasts 2--3 years.  Our results for the
Axelrod model may provide a hint toward understanding how widely separated
time scales arise in these types of complex dynamical systems.

\begin{figure}[ht]
 \vspace*{0.cm}
 \includegraphics*[width=0.40\textwidth]{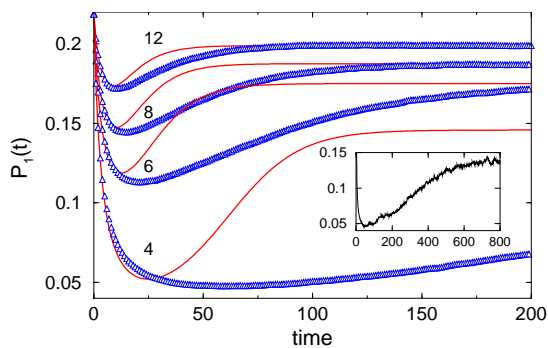}
 \caption{Active bond density from the master equations (curves) and from
   simulations of $10^2$ realizations ($\triangle$) on a degree-regular
   random graph with $10^4$ nodes for various coordination numbers, and $q=8$
   states per feature.  Inset: One realization with coordination number 4.}
\label{me-rg}
\end{figure}

Following Refs.~\cite{CMV,VVC}, we describe the Axelrod model in a minimalist
way by the density $P_m$ of bonds of type $m$.  These are bonds between
interaction partners in which there are $m$ common features.  This
description is convenient for monitoring the activity level in the system and
has the advantage of being analytically tractable.  We consider a mean-field
system in which each agent can interact with a fixed number of
randomly-selected agents.  Agents can thus be viewed as existing on the nodes
of a degree-regular random graph.  Such a topology is an appropriate setting
for cultural interaction, where both geographically nearby and distant
individuals may interact with equal facility.  We verified that simulations
of the Axelrod model on degree-regular random graphs qualitatively agree with
our analytical predictions, and this agreement becomes progressively more
accurate as the number of neighbors increases (Fig.~\ref{me-rg}).  Thus the
master equation approach describes the Axelrod model when random connections
between agents exist.


\begin{figure}[ht]
 \vspace*{0.cm}
 \includegraphics*[width=0.25\textwidth]{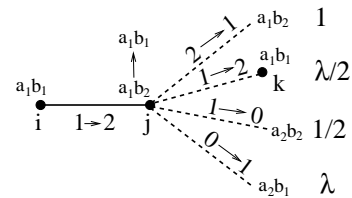}
 \caption{Illustration of the state-changing bond updates when agent $j$
   changes state from $a_1b_2\to a_1b_1$.  The values at the right give the
   relative rates of each type of event.}
\label{states}
\end{figure}

If interaction partners share no common features ($m=0$) or if all features
are common ($m=F$), then no interaction occurs across the intervening bond.
Otherwise, two agents that are connected by an active bond of type $m$ (with
$0<m<F$) interact with probability $m/F$, after which the bond necessarily
becomes type $m+1$.  In addition, when an agent changes a preference, the
index of all bonds attached to this agent may either increase or decrease
(Fig.~\ref{states}).  The competition between these direct and indirect
interaction channels underlies the rich dynamics of the Axelrod model.

Because we obtain qualitatively similar behavior for the density of active
links, $P_a\equiv \sum_{k=1}^{F-1} P_k$, for all $F \geq 2$, we focus on the
simplest non-trivial case of $F=2$.  For this example, there are three types
of bonds: bonds of type $0$ (no shared features) and type $2$ (all features
shared) are inert, while bonds of type $1$ are active.  As $q\to q_c$ from
below, $P_1$ is non-monotonic, with an increasingly deep minimum
(Fig.~\ref{P1vst}), while for $q>q_c$, $P_1$ decays to zero exponentially
with time.  There is a discontinuous transition at $q_c$ from a stationary
phase where the steady-state density of active links $P_a^s$ is greater than
zero to a frozen phase where $P_1^s=0$.

When fluctuations are neglected, the evolution of the bond densities $P_m$
when a single agent changes its state is described by the master equations:
\begin{eqnarray}
\label{eqn-P0}
\frac{dP_0}{dt}&\!=\!&\frac{\eta}{\eta\!+\!1} P_1 \left[ -\lambda P_0 + 
\frac{1}{2} P_1 \right]\,, \\
\label{eqn-P1}
\frac{dP_1}{dt}&\!=\!&\!-\frac{P_1}{\eta\!+\!1}+\frac{\eta}{\eta\!+\!1} P_1\! 
\left[ \lambda P_0 - 
                             \frac{1\!+\!\lambda}{2} P_1 + P_2 \right], \\
\label{eqn-P2}
\frac{dP_2}{dt}&\!=\!&\frac{P_1}{\eta\!+\!1}+\frac{\eta}{\eta\!+\!1} P_1 
\left[ \frac{\lambda}{2} 
 P_1 - P_2 \right]\,,
\end{eqnarray}
where $\eta+1$ is the network coordination number.  The first term on the
right-hand sides of Eqs.~\eqref{eqn-P1} and \eqref{eqn-P2} account for the
direct interaction between agents $i$ and $j$ that changes a bond of type $1$
to type $2$.  For example, in the equation for $\frac{dP_1}{dt}$, a type-1
bond and the shared feature across this bond is chosen with probability
$P_1/2$ in an update event.  This update decrements the number of type-1
bonds by one in a time $dt=\frac{1}{N}$, where $N$ is the total number of
sites in the system.  Assembling these factors gives the term
$-\frac{P_1}{\eta+1}$ in Eq.~\eqref{eqn-P1}.

The remaining terms in the master equations represent indirect interactions.
For example, if agent $j$ changes from $(a_1,b_2)$ to $(a_1,b_1)$ then the
bond to agent $k$ in state $(a_1,b_1)$ changes from type 1 to type 2
(Fig.~\ref{states}).  The probability for this event is proportional to
$P_1\lambda/2$: $P_1$ accounts for the probability that the indirect bond is
of type 1, the factor 1/2 accounts for the fact that only the first feature
of agents $j$ and $k$ can be shared, while $\lambda$ is the conditional
probability that $i$ and $k$ share one feature that is simultaneously not
shared with $j$.  If the distribution of preferences is uniform, then
$\lambda = (q-1)^{-1}$.  As the system evolves $\lambda$ generally depends on
the densities $P_m$.  Here we make an assumption of a mean-field spirit that
$\lambda$ stays constant during the dynamics \cite{VVC}; this makes the
master equations tractable.  Our simulations for random graphs with large
coordination number match the master equation predictions and give $\lambda$
nearly constant and close to $(q-1)^{-1}$ (Fig.~\ref{me-rg}), thus justifying
the assumption.

Let us first determine the stationary solutions of the master equations.  A
trivial steady state is $P_1^s =0$, corresponding to a static society.  A
more interesting stationary solution is $P_1^s>0$, corresponding to
continuous evolution; as we shall see, this dynamic state arises when
$q<q_c$.  Setting $\frac{dP_i}{dt}=0$ in the master equations and solving, we
obtain:
\begin{eqnarray}
\label{Pss}
P_0^s&=&\frac{(\eta -1)}{\eta (1+\lambda)^2}\,, \quad
P_1^s=\frac{2 \lambda (\eta -1)}{\eta (1+\lambda)^2}\,, \nonumber \\
P_2^s&=&\frac{(1+\lambda)^2+\lambda^2(\eta -1)}{\eta (1+\lambda)^2}.
\end{eqnarray} 

Since $\lambda=\lambda(q)$ is the only parameter in the master equations, the
two stationary solutions suggest that there is a transition at a critical
value $q_c$ such that both solutions apply, but on different sides of the
transition.  To locate the transition, it proves useful to relate $P_1$ and
$P_2$ directly.  Thus we divide Eq.~(\ref{eqn-P1}) by Eq.~(\ref{eqn-P2}) and
eliminate $P_0$ via $P_0=1-P_1-P_2$ and obtain, after some algebra:
\begin{equation}
\frac{dP_1}{dP_2}=\frac{-1+\eta \lambda-\frac{1}{2}\eta (1+3\lambda)P_1+\eta(1-\lambda)P_2} 
                      {1+\frac{1}{2}\eta \lambda P_1-\eta P_2}\,.
\label{dP1dP2}
\end{equation}

The solution to Eq.~\eqref{dP1dP2} has the form
\begin{eqnarray}
\label{trial}
P_1=\alpha+\beta P_2-\sqrt{\gamma+\delta P_2}\,,
\end{eqnarray} 
where we determine the coefficients $\alpha$, $\beta$, $\gamma$ and $\delta$
by matching terms of the same order in Eq.~\eqref{dP1dP2} and in
$\frac{dP_1}{dP_2}$ from Eq.~\eqref{trial}.  The procedure gives the solution
except for one constant that is specified by the initial conditions.  For the
initial condition where features for each agent are chosen uniformly from the
integers $[0,q-1]$, the distribution of initial bond densities is binomial,
$P_m(t=0)=\frac{2!}{m!  (2-m)!}(1/q)^m(1-1/q)^{2-m}$.  Matching this initial
condition to the solution of Eq.~\eqref{trial} gives:
\begin{eqnarray}
P_1(P_2)& = &\frac{2\lambda}{1+\lambda}+\frac{2}{\eta}-2P_2 \nonumber \\
&~&~~~~-\frac{2}{\eta}\frac{\sqrt{\eta\lambda^2+(1+\lambda)^2(1-\eta P_2)}}{(1+\lambda)}\,.
\label{P1-P2}
\end{eqnarray}

\begin{figure}[ht]
 \vspace*{0.cm}
 \includegraphics*[width=0.425\textwidth]{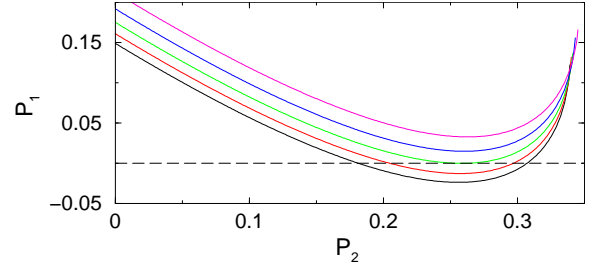}\hskip 0.1cm
 \caption{$P_1$ vs $P_2$ from Eq.~(\ref{P1-P2}) for $\eta=3$ and $q=q_c-2$,
   $q_c-1$, $q_c$, $q_c+1$ and $q_c+2$ (top to bottom).}
\label{P1vsP2}
\end{figure}

As a function of $P_2$, $P_1$ has a minimum $P_1^{\rm min}(q)$ that
monotonically decreases as $q$ increases and becomes negative for $q$ larger
than a critical value $q_c$ (Fig.~\ref{P1vsP2}).  The phase transition
between the active and the frozen state corresponds to the value of $q$ where
$P_1$ first reaches zero.  To find $q_c$, we calculate $P_1^{\rm min}$ as a
function of $\lambda(q)$ from Eq.~(\ref{P1-P2}) and then find the value of
$q$ at which $P_1^{\rm min}$ becomes zero.  This leads to
\begin{eqnarray*}
P_1^{\rm min}=\frac{4 \eta \lambda-(1+\lambda)^2}{2\eta(1+\lambda)^2}
\equiv \frac{S(\lambda,\eta)}{2\eta(1+\lambda)^2} \,, 
\end{eqnarray*}
from which the critical point is given by
\begin{eqnarray*}
q_c&=&2\eta+2\sqrt{\eta(\eta-1)}\,,
\end{eqnarray*} 
while $P_1^{\rm min} \propto S\propto (q_c-q)$ for $q<q_c$.

We now determine the steady-state bond densities in the frozen state.  From
Eq.~\eqref{P1-P2}, we compute the stationary value $P_2^s$ at the point where
$P_1$ first reaches zero.  The smallest root of this equation then gives
\begin{eqnarray*}
P_2^s=\frac{1+\lambda+2\eta \lambda-\sqrt{(1+\lambda)^2-4 \eta \lambda}}
           {2 \eta (1+\lambda)}\,,
\end{eqnarray*}
while $P_0^s=1-P_2^s$.

The most interesting behavior is the time dependence of the density of active
bonds, $P_1(t)$.  We solve for $P_1(t)$ by first inverting Eq.~\eqref{P1-P2}
to express $P_2$ in terms of $P_1$
\begin{eqnarray*}
P_2(P_1)=\frac{1+\lambda(1+2\eta)}{2\eta(1+\lambda)}-\frac{P_1}{2}-
\frac{\sqrt{2\eta(1+\lambda)^2P_1-S}}{2\eta(1+\lambda)}\,,
\label{P2-P1}
\end{eqnarray*}
and then writing $P_0=1-P_1-P_2(P_1)$ also in terms of $P_1$, and finally
substituting these results into the master equation \eqref{eqn-P1} for $P_1$.
After some algebra, we obtain
\begin{eqnarray}
\frac{dP_1}{d\tau}&=&S P_1-(1-\lambda)\sqrt{2\eta (1+\lambda)^2 P_1-S\,}\,P_1 \nonumber \\
                  &~&~~~~~~- 2\eta(1+\lambda)^2P_1^2,
\label{dP1dtau}
\end{eqnarray}
where we use the rescaled time variable $\tau=
\frac{t}{2(\eta+1)(1+\lambda)}$.  This master equation can be simplified by
substituting the quantity $\Delta\equiv 2\eta(1+\lambda)^2P_1-S$, which
measures the deviation of $P_1$ from its minimum value, in
Eq.~(\ref{dP1dtau}).  We obtain
\begin{equation}
\frac{d\Delta}{d\tau}=-\sqrt{\Delta}(S+\Delta)(1-\lambda+\sqrt{\Delta})\,.
\label{dDdtau}
\end{equation}
Performing this integral by partial fraction expansion gives
\begin{eqnarray}
\label{tau-P1}
\tau=\frac{1}{4\lambda(\eta-1)} \Biggl[ \ln \left( \frac{S+\Delta}
      {\eta\lambda(1-\lambda)^2} \right)-
     2 \ln \left( 1 \pm \frac{\sqrt{\Delta}}{1-\lambda} \right) \nonumber \\
      + \frac{1-\lambda}{\sqrt{-S}} 
  \ln \left( \frac{(\sqrt{-S}-1-\lambda)(\sqrt{-S}\pm \sqrt{\Delta})}
        {(\sqrt{-S}+1+\lambda)(\sqrt{-S}\mp \sqrt{\Delta})}\right) \Biggr]
      \,. \nonumber \\ 
\end{eqnarray}
For $q>q_c$, only the upper sign is needed.  For $q<q_c$, the upper sign
applies for $t<t^{\rm min}$ and the lower sign applies for $t>t^{\rm min}$;
here $t^{\rm min}$ is the time at which $P_1(t)$ reaches its minimum value.
Substituting back $t$ and $P_1$ in Eq.~\eqref{tau-P1} gives the formal exact
solution of Eq.~\eqref{dP1dtau}.

For $q<q_c$, we determine $P_1(t)$ near its minimum by taking the $\Delta\to
0$ limit of Eq.~\eqref{dDdtau}.  This gives
\begin{equation}
  \frac{d\Delta}{dt}\approx -aS\sqrt{\Delta}\,,
\label{dedt<}
\end{equation}
with $a=\frac{(1-\lambda)}{2(\eta+1)(1+\lambda)}>0$.  For $S>0$, the solution
to the lowest-order approximation shows that $P_1$ has a quadratic form
around its minimum:
\begin{eqnarray}
P_1(t)-P_1^{\rm min} \propto \Delta \approx 
\frac{a^2S^2}{8\eta(1+\lambda)^2}(t-t_{\rm min})^2\,.
\end{eqnarray} 
When $q\to q_c$, the factor $S$ may be neglected as long as $\Delta>S$, and
this leads to $\Delta$ decaying as $t^{-2}$ before the minimum in $P_1$ is
reached (dashed line in Fig.~\ref{P1vst}).

The peculiar behavior of $P_1$ as a function of time for $q$ below but close
to the critical value $q_c$ is shown in Fig.~\ref{P1vst}.  The density of
active bonds quickly decreases with time and this decrease extends over a
wide range when $q$ is close to $q_c$.  Thus on a linear scale, $P_1$ remains
close to zero for a substantial time.  After a minimum value at $t_{\rm min}$
is reached, $P_1$ then increases and ultimately reaches a non-zero asymptotic
value for $q<q_c$.  The quasi-stationary regime where $P_1$ remains small is
defined by: (i) a time scale of the order of one that characterizes the
initial decay of $P_1(t)$, and (ii) a much longer time scale $t_{\rm asymp}$
where $P_1$ rapidly increases and then saturates at its steady-state value.

We can give a partial explanation for the time dependence of $P_1$.  For
$q>q_c$, there are initially small enclaves of interacting agents in a frozen
discordant background.  Once these enclaves reach local consensus, they are
incompatible with the background and the system freezes.  For $q\alt q_c$
(less diversity), sufficient active interfaces are present to slowly and
partially coarsen the system into tortuous domains whose occupants are either
compatible (that is, interacting) or identical.  Within a domain of
interacting agents, the active interface can ultimately migrate to the domain
boundary and facilitate merging with other domains; this corresponds to the
sharp drop in $P_0$ seen in Fig.~\ref{Pm} \cite{applet}. While this picture
is presented in the context of a lattice system, it remarkably still seems to
apply for degree-regular random graphs and in a mean-field description.

Both $t_{\rm min}$ as well as the end time of the quasi-stationary period
$t_{\rm asymp}$ increase continuously and diverge as $q$ approaches $q_c$
from below.  To find these divergences, we expand $t_{\rm min}$ and $t_{\rm
  asymp}$ in powers of $S$.  From Eq.~(\ref{tau-P1}), the first two terms in
the expansion of $t_{\rm min}$, as $S\to 0$, are
\begin{eqnarray*}
t_{\rm min}=t(P_1^{\rm min}) \approx A\ln S+\frac{B}{\sqrt{S}} \sim 
                 \frac{B}{\sqrt{S}}\,,
\end{eqnarray*}
where $A,B$ are constants.  As a result, $t_{\rm min} \sim (q_c-q)^{-1/2}$ as
$q \to q_c$.  Similarly, we estimate $t_{\rm asymp}$ as the time at which
$P_1$ reaches one-half of its steady-state value.  Using Eqs.~\eqref{Pss} and
\eqref{tau-P1}, we find
\begin{eqnarray*}
t_{\rm asymp}=t(P_1^s/2) \sim \frac{1}{\sqrt{S}}\;\;\mbox{as} \;\;S \to 0\,,
\end{eqnarray*}
so that $t_{\rm asymp} \sim (q_c-q)^{-1/2}$ as $q \to q_c$.

For $q>q_c$, the system evolves to a frozen state with $P_1\to 0$.  To lowest
order Eq.~\eqref{dP1dtau} becomes $\frac{dP_1}{dt}=-\frac{P_1}{\mathcal{T}}$,
with $\mathcal{T}=\frac{2(\eta+1) (1+\lambda)}{-S + (1-\lambda)\sqrt{-S}}$.
Here $\mathcal{T}>0$ since $S < 0$ for $q>q_c$.  Consequently $P_1$ decays
exponentially in time as $t \to \infty$.  As $q$ approaches $q_c$, $S$
asymptotically vanishes as $(q_c-q)$ and the leading behavior is $\mathcal{T}
\sim (q-q_c)^{-1/2}\,$.  Thus again there is an extremely slow approach to
the asymptotic state as $q$ approaches $q_c$.

In summary, the density of active links is non-monotonic in time and is
governed by an anomalously long time scale in the 2-feature and q preferences
per feature Axelrod model.  For $q<q_c$, an active steady-state state is
reached in a time that diverges as $(q_c-q)^{-1/2}$ when $q\to q_c$ from
below.  For $q>q_c$, the final state is static and the time scale to reach
this state also diverges as $(q_c-q)^{-1/2}$ as $q\to q_c$ from above.


We gratefully acknowledge financial support from the US National Science
Foundation grant DMR0535503.

\end{document}